\documentclass[reprint,aps,prc,twocolumn,superscriptaddress,floatfix,amsmath,amssymb,10pt]{revtex4-2}
\DeclareMathAlphabet{\mathpzc}{OT1}{pzc}{m}{it}
\usepackage{bm}
\usepackage{dcolumn}
\usepackage{amsthm}
\usepackage{amsmath}
\usepackage{amssymb}
\usepackage{graphicx}
\usepackage{xcolor}
\usepackage{fix-cm}
\usepackage{mathptmx} 
\usepackage[T1]{fontenc}
\usepackage[colorlinks,allcolors=blue]{hyperref}
\makeatletter
\def\NAT@def@citea{\def\@citea{\NAT@separator}}
\makeatother

\begin{document}

\title{Cluster model of $^{12}$C in density functional theory framework}

\author{A.S. Umar}\email{umar@compsci.cas.vanderbilt.edu}
\affiliation{Department of Physics and Astronomy, Vanderbilt University, Nashville, Tennessee 37235, USA}
\author{K. Godbey}\email{godbey@frib.msu.edu}
\affiliation{Facility for Rare Isotope Beams, Michigan State University, East Lansing, Michigan 48824, USA }
\author{C. Simenel}\email{cedric.simenel@anu.edu.au}
\affiliation{Department of Fundamental and Theoretical Physics and Department of Nuclear Physics and Accelerator Applications, Research School of Physics, The Australian National University, Canberra ACT 2601, Australia}
\date{\today}


\begin{abstract}
We employ the constrained density functional theory to investigate cluster phenomena for
the \textsuperscript{12}C nucleus. The proton and neutron densities are generated from the placement
of three \textsuperscript{4}He nuclei (alpha particles) geometrically. These densities are then used in a
density constrained Hartree-Fock calculation that produces an antisymmetrized state with the same densities through energy minimization.
In the calculations no \textit{a priori} analytic form for the
single-particle states is assumed and the full energy density functional is utilized.
The geometrical scan of the energy landscape provides the ground state of
\textsuperscript{12}C as an equilateral triangular configuration of three alphas with
molecular bond like structures.
The use of the nucleon localization function provides further
insight to these configurations.
One can conclude
that these configurations are a hybrid between a pure mean-field and a pure alpha particle
condensate. This development could facilitate DFT based fusion calculations with a more realistic
\textsuperscript{12}C ground state.
\end{abstract}
\maketitle


\section{Introduction}
 In stellar evolution carbon plays a pivotal role through the
carbon burning process. The ignition of carbon burning for
stars in the mass range M$>$8-10 solar masses lead to white
Ne/O dwarfs, while massive stars with masses M$>$25 solar masses
can continue burning Ne, O, and Si and end up as supernovae.
Similarly, Type Ia supernovae is believed to result from
an explosion of a white dwarf accreting mass from
a binary companion or a merger, inducing high enough temperatures to ignite
carbon in the core leading to a supernovae explosion.
Superbursts are set off by the ignition of carbon in the accumulated
ashes of previous x-ray bursts~\cite{barnes1985,hansen2003,toonen2012}.
Overall, a change in the \textsuperscript{12}C+\textsuperscript{12}C reaction rates has a profound
impact on all these mechanisms as well as nucleosynthesis~\cite{fruet2020,mori2018,tumino2018,cooper2009,spillane2007,tan2020,monpribat2022,adsley2022}.
In the stellar environment carbon is produced in a two step process;
in the first step two \textsuperscript{4}He nuclei come together to form an unstable
\textsuperscript{8}Be, which decays back to two \textsuperscript{4}He nuclei
with a very short lifetime ($10^{-16}$s).
However, during the helium burning stage, the densities are high enough to maintain
a small abundance of \textsuperscript{8}Be, which renders it possible to combine
with another \textsuperscript{4}He to form an excited carbon nucleus
through the well known Hoyle state of \textsuperscript{12}C. With a much lower
probability a triple \textsuperscript{4}He combination may also lead to the
same outcome.

In addition to its astrophysical importance, the microscopic
description of the carbon nucleus, which is an essential ingredient to
the reaction calculations, has proven to be a challenge.
This is predicated by the expectation that the structure of carbon
should exhibit a pronounced cluster structure.
Clustering effects are widely believed to play prominent role
in the structure of N$=$Z nuclei, resulting in a molecular type
phenomenon. To what degree such nuclei can be viewed as being
comprised of a pure alpha particle condensate is still an open
question~\cite{freer2018}.

However, employing the standard non-relativistic density functional theory (DFT) results
in a ground state of \textsuperscript{12}C without any sign of clusterization
~\cite{reinhard2011}. This is also true
for \textsuperscript{16}O, while the ground states of \textsuperscript{8}Be and \textsuperscript{20}Ne do show some cluster features~\cite{reinhard2011}.
Relativistic mean-field theories seem to favor more clusterization due to deeper
potentials~\cite{ebran2012}.
Alternate approach of using configuration mixing with generator coordinate method (GCM) calculations using the Skyrme energy
density functional have also been done in Refs.~\cite{shinohara2006,fukuoka2013}.
Furthermore, these calculations using modern energy density
functionals commonly result in a spherical ground state for the \textsuperscript{12}C nucleus, which
is experimentally known to have an oblate deformation~\cite{yasue1983,raju2018,marinlambarri2014}.
On the other hand the excited states of these nuclei do seem
to exhibit some cluster structure, e.g. the linear-chain configuration
of \textsuperscript{12}C, which originally was thought to be the Hoyle state~\cite{morinaga1956}, and
excited states obtained by various constraints~\cite{tohsaki2001,ichikawa2011,ebran2012,girod2013,ebran2014,funaki2015,zhao2015b,ebran2017,marevic2019,ichikawa2022,wang2022}.
Such formations are also observed via the time-dependent
Hartree-Fock studies of the triple-alpha reaction~\cite{umar2010c,stevenson2020b}
and studied in recent experiments~\cite{bishop2020,bishop2022}.
In these time-dependent calculations a bent-arm intermediate configuration is observed
during the decay of the metastable linear chain state of \textsuperscript{12}C.

The fact that DFT calculations account for only limited clustering effects prompted alternate approaches
to study \textsuperscript{12}C structure that rely more heavily on cluster wavefunctions.
These calculations suggest that a substantial contribution of alpha cluster correlations
that are not accounted for in the mean-field description should nevertheless be present in \textsuperscript{12}C states.
These include various \textit{ab initio}
calculations~\cite{epelbaum2011,epelbaum2012,shen2022} as well as approaches that are
collectively referred to as
molecular dynamics that significantly extend the original Bloch-Brink
alpha cluster model~\cite{brink1966}.
In Brink's approach each quartet of
nucleons where represented using harmonic oscillator wavefunctions
with zero angular momentum displaced from each other by a relative
coordinate. Antisymmetrization followed by normalization comprised
the many-body wavefunction in terms of the locations of the quartets.
These quartets, interacting via an effective nucleon-nucleon interaction,
are optimized with respect to their size and position to map out
the energy landscape showing the location of the minima. This
approach was extended via the resonating group method as well
as generator coordinate method to better incorporate the internal
structure of the clusters. The antisymmetrized molecular dynamics (AMD)~\cite{kanada1995,kanada2007,kanada2022}
approach employs a Slater determinental many-body wavefunction
comprised of single-particle states as Gaussian wave-packets
using an advanced set of geometrical variables.
Fermionic molecular dynamics (FMD)~\cite{neff2004} further extends AMD by not putting
any restriction on the width of these Gaussians.
These calculations indicate a large admixture of alpha-cluster triangular states
for the ground and some of the excited states configurations of \textsuperscript{12}C.
The use of Gaussian basis and suitable interactions allow for
very powerful extensions for the above methods, such as the
treatment of the center-of-mass energy, angular momentum
projection and the use of the generator coordinate method (GCM).
However, many of these calculations assume a degeneracy between neutron and proton wave functions
and do not include the full effective interaction and the spin-orbit force.
A collection of recent reviews can be found in Refs.~\cite{vonoertzen2006,funaki2008,freer2007,yoshida2016,tohsaki2017,freer2018}.

In this manuscript we introduce another approach for studying
cluster structures within the DFT framework. This is accomplished
through the use of the density constrained Hartree-Fock approach.
Here, we start with alpha-particles as solutions to the unconstrained
Hartree-Fock (HF) equations. These alpha particles are then geometrically
arranged on the numerical grid defining the total density of the system.
For each arrangement, a mean-field solution is obtained through minimization of
the energy by constraining the density of the entire system.
Density constraint iterations allow for the rearrangement of the
single-particle states through their orthogonalization and energy
minimization. This takes care of antisymmetrization as well as the
overall energy dependent normalization of the many-body wavefunction.
No assumption about the mathematical form of the single-particle states is made
and the full effective interaction, including the Coulomb force, can be used.
We also employ the nuclear localization function (NLF),
which  allows for a more precise characterization of spatial distributions.
This method blends the cluster based approach with the fully microscopic approach.
As we shall see, it has advantages and certain disadvantages.

\section{Microscopic methods}\label{sec1}
In this section we briefly outline the formalisms and methods used in our calculations. Further details can be found
in the cited references.

\subsection{Density constraint}
Given a reference density, the density constraint procedure~\cite{cusson1985,umar1985}
allows the single-particle states,
comprising the combined nuclear density, to reorganize
to attain their minimum energy configuration and be properly antisymmetrized as the many-body
state is a Slater determinant of all the occupied single-particle wave-functions.
Here, the reference density is given by the combined density of three alpha particles obtained from independent
Hartree-Fock calculations and placed in close proximity of each other.
The HF minimization of the combined system is thus performed subject to the constraint that the
local proton ($p$) and neutron ($n$) densities do not change:
\begin{equation}
    \delta \left\langle \ H -\sum_{q=p,n}\int\, d\mathbf{r} \ \lambda_q(\mathbf{r})  \left[ \sum_{i=1}^{3}\rho^{\alpha}_{i_q}(\mathbf{r},\mathbf{R_i})\right] \ \right\rangle = 0\,,
    \label{eq:var_dens}
\end{equation}
where the $\lambda_{n,p}(\mathbf{r})$ are Lagrange parameters at each point
of space constraining the neutron and proton local densities, $\rho^{\alpha}_{i_q}(\mathbf{r},\mathbf{R_i})$
is the proton/neutron densities of an alpha particle located at position $\mathbf{R_i}$, and $H$ is the
effective many-body Hamiltonian.
This procedure determines a unique Slater determinant $|\Phi(\mathbf{R_1},\mathbf{R_2},\mathbf{R_3})\rangle$
for the combined system.
The density constraint has been extensively used in the calculation of ion-ion interaction
barriers for fusion calculations~\cite{simenel2018,simenel2017,umar2021}.

\subsection{Centre of mass correction}
A major drawback of any mean-field based microscopic calculation is the uncontrolled
presence of the energy associated with the center-of-mass (c.m.) motion~\cite{umar2009c}. This energy
is particularly large for light nuclei. Most Skyrme interactions adopt a simple one-body correction
for this energy, which may be reasonable for heavy systems. This issue has been discussed
more extensively in the context of alpha clustering phenomenon for the mean-field calculations
in Ref.~\cite{girod2013}, where a constant value of 7~MeV per alpha particle was subtracted.
However, the c.m. correction for a composite system may not be the same as adding corrections for each alpha.
Due to this we cannot make any binding energy comparisons and adopted the SLy4d interaction, which
does not employ any center-of-mass correction term.

\subsection{The nucleon localization function (NLF)}
The measure of localization has been originally developed in the context of
a mean-field description for electronic systems~\cite{becke1990},
and subsequently introduced to nuclear systems~\cite{reinhard2011,li2020,khan2022}.
We first realize that a fermionic mean-field state is fully characterized by the one-body density-matrix
$\rho_q(\mathbf{r}s,\mathbf{r'}s')$.
The probability of finding two nucleons with the same spin at spatial
locations $\mathbf{r}$ and $\mathbf{r'}$ (same-spin pair probability) for
isospin $q$ is proportional to
\begin{equation}
    P_{qs}(\mathbf{r},\mathbf{r}') =
    \rho_q(\mathbf{r}s, \mathbf{r}s)\rho_q(\mathbf{r}'s, \mathbf{r}'s)
    -
    |\rho_q(\mathbf{r}s,\mathbf{r}'s)|^2\,,
\end{equation}
which vanishes for $\mathbf{r}=\mathbf{r'}$ due to the Pauli exclusion principle.
The conditional probability for finding a nucleon at
$\mathbf{r'}$ when we know with certainty that another nucleon with the same
spin and isospin is at $\mathbf{r}$ is proportional to
\begin{equation}
    \label{conditionalProb}
    R_{qs}(\mathbf{r},\mathbf{r}')
    =
    \frac{P_{qs}(\mathbf{r},\mathbf{r}')}{\rho_q(\mathbf{r}s,\mathbf{r}s)}\,.
\end{equation}
The short-range behavior of $R_{qs}$ can be
obtained using techniques similar to the local density approximation~\cite{reinhard2011,li2020}. The
leading term in the expansion yields the localization measure
\begin{equation}\label{eq:prob_D}
    D_{qs_{\mu}} = \tau_{qs_{\mu}}-\frac{1}{4}\frac{\left|\boldsymbol{\nabla}\rho_{qs_{\mu}}\right|^2}{\rho_{qs_{\mu}}}
    -\frac{\left|\mathbf{j}_{qs_{\mu}}\right|^2}{\rho_{qs_{\mu}}}\,.
\end{equation}
This measure is the most general form that is appropriate for deformed nuclei and without assuming
time-reversal invariance, thus also including the time-odd terms important in applications such as
cranking or time-dependent Hartree-Fock (TDHF). The densities and currents are given in their most unrestricted  form~\cite{engel1975,bender2003,perlinska2004}
for $\mu$-axis denoting the spin-quantization axis by~\cite{li2020}
\begin{subequations}\label{eq:density_relation}
    \begin{align}
        \rho_{q s_\mu}(\mathbf{r}) &= \frac{1}{2}\rho_q(\mathbf{r}) + \frac{1}{2}\sigma_\mu {s}_{q\mu}(\mathbf{r})\,, \\
        \tau_{q s_\mu}(\mathbf{r}) &= \frac{1}{2}\tau_q(\mathbf{r}) + \frac{1}{2}\sigma_\mu {T}_{q\mu}(\mathbf{r})\,, \\
        \mathbf{j}_{q s_\mu}(\mathbf{r}) &= \frac{1}{2}\mathbf{j}_q(\mathbf{r}) + \frac{1}{2}\sigma_\mu \mathbb{J}_q(\mathbf{r})\cdot \mathbf{e}_\mu\,,
    \end{align}
\end{subequations}
where $\sigma_\mu =2s_\mu=\pm1$ and $\mathbf{e}_\mu$ is the unit vector in the direction of the $\mu$-axis.
Note that subscripts $s_{\mu}$ denote spin along the quantization axis and should not be confused by the
spin-density ${s}_{q\mu}$.
The dot product in Eq.~(\ref{eq:density_relation}c) is explicitly given in the case of \textit{e.g.}, $\mu=z$
\begin{equation*}
    \mathbb{J}_q(\mathbf{r})\cdot \mathbf{e}_z=\frac{1}{2i}\left[ (\boldsymbol{\nabla}-\boldsymbol{\nabla}')
    s_{qz}(\mathbf{r},\mathbf{r}')\right]_{\mathbf{r}=\mathbf{r}'}\;.
\end{equation*}
The explicit expressions of the local densities and currents are given in Refs.~\cite{engel1975,li2020}.
We note that the localization measure includes the spin-density ${s}_{q\mu}(\mathbf{r})$, the time-odd part of the kinetic density ${T}_{q\mu}(\mathbf{r})$,
as well as the full spin-orbit tensor $\mathbb{J}_q(\mathbf{r})$, which is a pseudotensor. In this sense all of the terms in the
Skyrme energy density functional~\cite{engel1975} contribute to the measure.
Finally, we note that the time-odd terms contained in the above definitions
(${s}_{q\mu}$, ${T}_{q\mu}$, and $\mathbf{j}_q$) are zero in static calculations of
even-even nuclei but the spin-tensor $\mathbb{J}_q$ is not. Therefore, $\mathbf{j}_{q s_\mu}$
is not zero in general.

It is  interesting to visualize the NLF as it is also defined from the localization measure in Eq.~(\ref{eq:prob_D}).
We first normalize the localization measure using~\cite{li2020}
\begin{equation}
    \mathpzc{D}_{qs_\mu}(\mathbf{r})=\frac{D_{qs_\mu}(\mathbf{r})}{\tau_{qs_\mu}^{\mathrm{TF}}(\mathbf{r})}\,,
\end{equation}
where the normalization $\tau_{qs_\mu}^{\mathrm{TF}}(\mathbf{r})=\frac{3}{5}\left(6\pi^2\right)^{2/3}\rho_{qs_\mu}^{5/3}(\mathbf{r})$ is the Thomas-Fermi kinetic density.
The NLF can then be represented  either by $1/\mathpzc{D}_{qs_\mu}$ or by
\begin{equation}
    {C}_{qs_\mu}(\mathbf{r})=\left[1+\mathpzc{D}_{qs_\mu}^{2}\right]^{-1}
    \label{eq:NLF}
\end{equation}
which is used here.
The advantage of the latter form is that it scales to be in the interval $[0,1]$, but otherwise both forms
show similar localization details.

The information content of the localization function is better understood by considering limiting cases.
The extreme case of ideal metallic bonding is
realized for homogeneous matter where $\tau=\tau_{q\sigma}^\mathrm{TF}$.  This
yields $C=\frac{1}{2}$, a value which thus signals a region with a
nearly homogeneous Fermi gas as it is typical for metal electrons, nuclear
matter, or neutron stars.
The opposite regime are space regions where exactly one single-particle
wavefunction of type $q\sigma$ contributes.  This is
called \textit{localization} in molecular physics.  Such a situation
yields $D_{q\sigma}(\mathbf{r})=0$,
since it is not possible to find another like-spin state in the vicinity,
and consequently $C=1$, the value which signals \textit{localization}.

In the nuclear case, it is the
$\alpha$ particle which is perfectly localized in this sense,
i.e. which has $C=1$ everywhere for all states.
Well bound nuclei show usually metallic bonding and predominantly have
$C=\frac{1}{2}$. Light nuclei are often expected to contain
pronounced $\alpha$-particle sub-structures.
Such a sub-structure means that in a certain region of space only
an $\alpha$ particle is found which in turn is
signaled by $C=1$ in this region.  In fact, an $\alpha$
sub-structure is a correlation of four particles: $p\uparrow$,
$p\downarrow$, $n\uparrow$, and $n\downarrow$. Thus it is signaled
only if we find simultaneously for all four corresponding localization
functions $C_{q\sigma}\approx 1$. This localization procedure was recently employed to
visualize the cluster structure in $N=Z$ light nuclei~\cite{matsumoto2022}.

\subsection{Numerical details}
Calculations were done in a three-dimensional
Cartesian geometry with no symmetry assumptions using
the code of Ref.~\cite{umar2006c} and using the
Skyrme SLy4d interaction~\cite{kim1997}, which has been successful in
describing various types of nuclear reactions~\cite{simenel2012,simenel2018}.
The three-dimensional Poisson equation for the Coulomb potential
is solved by using Fast-Fourier Transform techniques
and the Slater approximation is used for the Coulomb exchange term.
The static HF equations and the density constraint
minimizations are implemented using the damped gradient
iteration method~\cite{bottcher1989}. The box size used for all the calculations
was chosen to be $24\times 24\times 24$~fm$^3$, with a mesh spacing of
$1.0$~fm in all directions.
These values provide very accurate results due to the employment
of sophisticated discretization techniques~\cite{umar1991a,umar1991b}.

\section{Results}\label{sec2}
The placement of the three alpha particles were done as follows; two alpha particles were
placed on the $x-$axis with a spacing denoted by $d_2/2$ on each side of the origin. The
third alpha particle was placed at distance $z_3$ vertically from the origin.
There are numerous studies of alpha cluster models for $^{12}$C that show that this more
symmetric arrangement leads to the minimum energy configuration~\cite{yoshida2016}, as
anticipated from symmetry arguments. Moving the third alpha in the y-direction would simply
correspond to tilting the three-alpha system in the 3D space.
We have scanned $d_2$ values ranging from 1.5-6.6~fm in steps of 0.1~fm. For each value of $d_2$,
 $z_3$ was varied
from 0.7-5.0~fm in steps of 0.2~fm. When necessary we have used a smaller spacing to pinpoint
the desired location more precisely. For $z_3<0.7$ the large overlap among the three alphas
lead to convergence problems due to unphysically large densities.
\begin{figure}[!thb]
    \includegraphics[width=8.6cm]{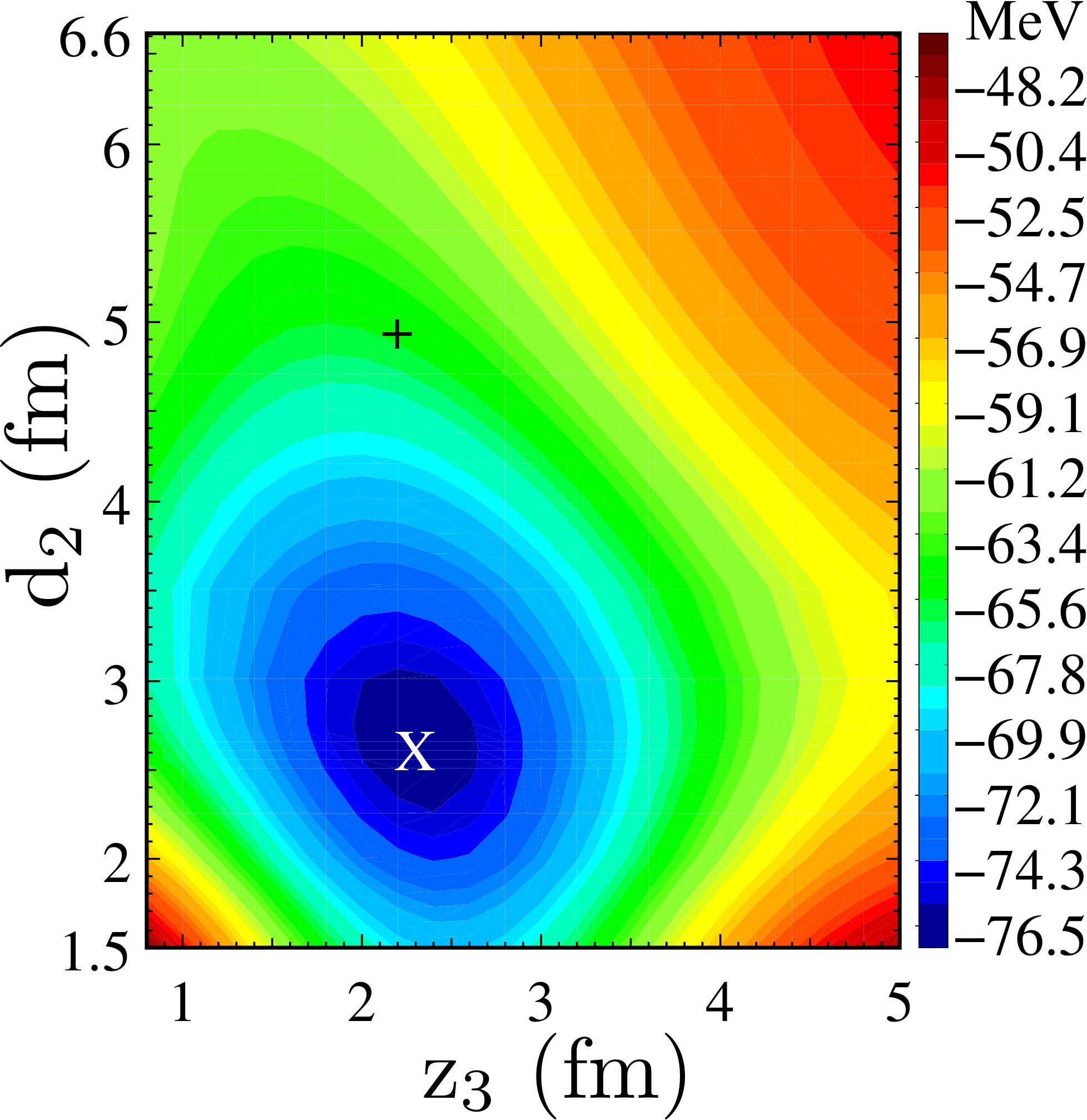}
    \caption{\protect The 3-$\alpha$ energy surface obtained from the density constraint procedure
        as a function of the spacings $d_2$ and $z_3$. The point marked by $X$ indicates the location
        of the minimum.}
    \label{fig:PES}
\end{figure}
\begin{figure}[!htb]
    \includegraphics[width=7.6cm]{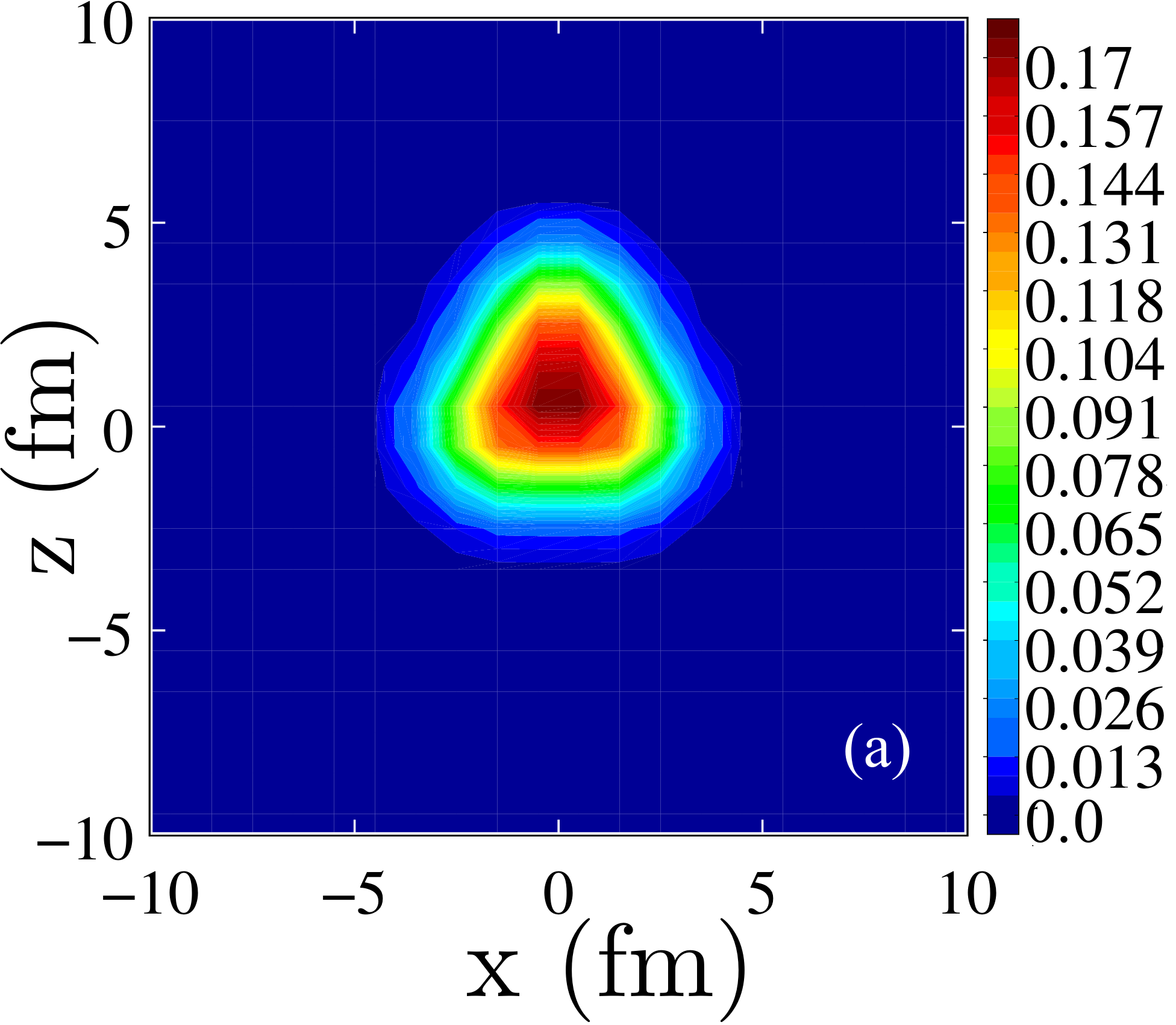}
    \includegraphics[width=7.6cm]{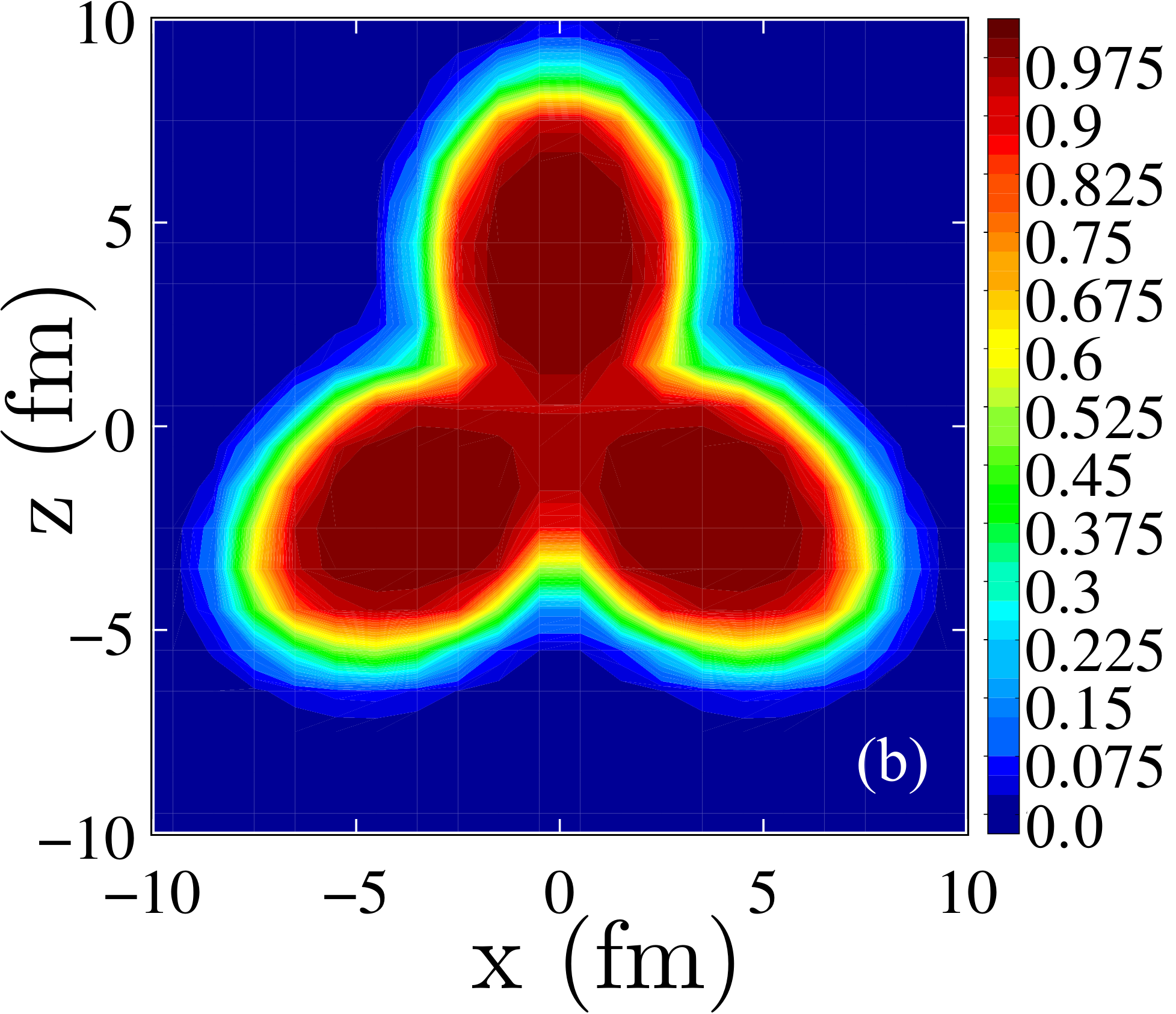}
    \includegraphics[width=7.6cm]{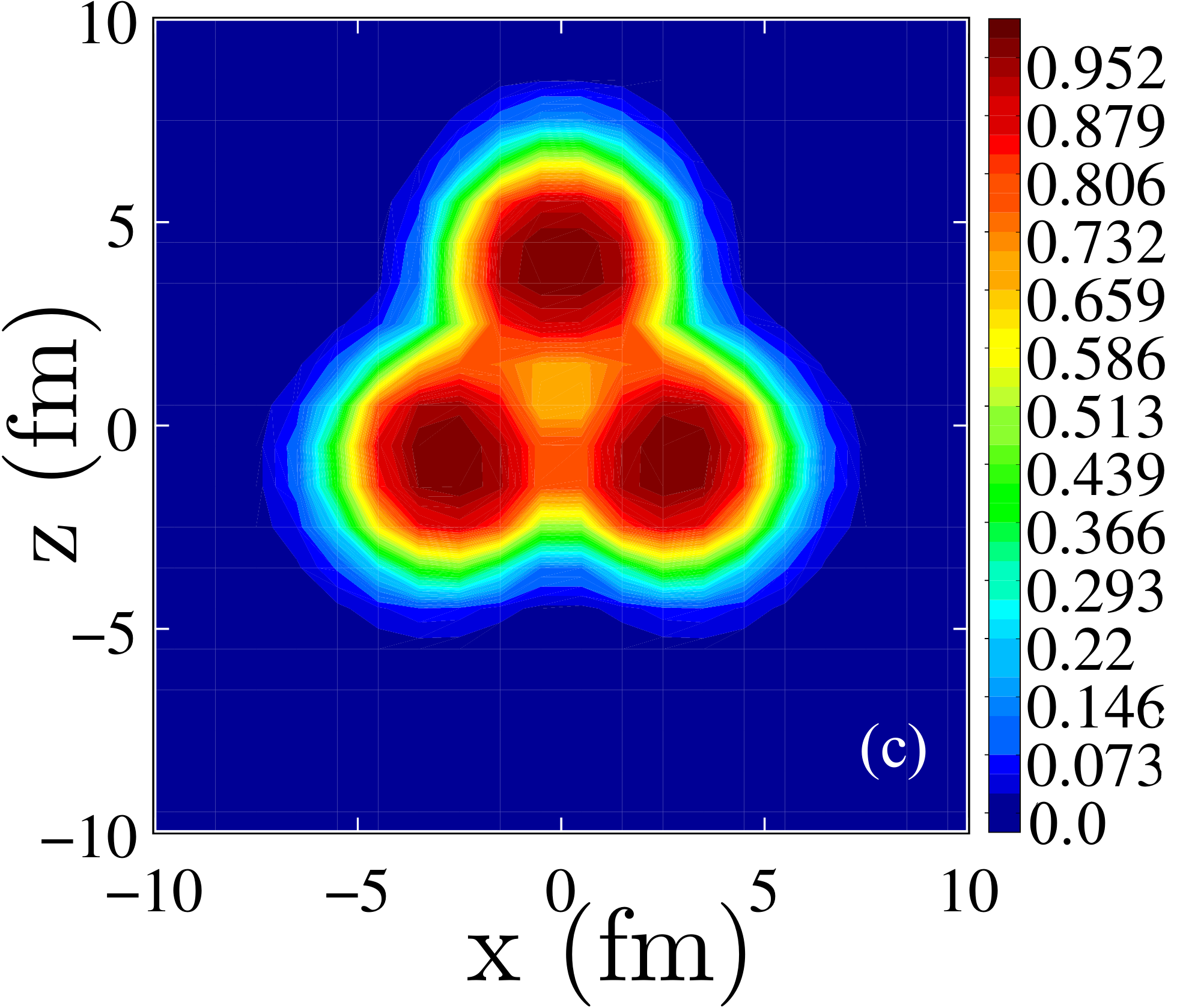}
    \caption{\protect
        (a) The total density for the ground state configuration of the three alpha particles plotted
        in the $x$-$z$ plane.
        (b) The $n\uparrow$ localization function before the density constraint.
        (c) The $n\uparrow$ localization function of the ground state configuration after the
        density constraint minimization.}
    \label{fig:gs}
\end{figure}

In Fig.~\ref{fig:PES} we plot the 3-$\alpha$  energy surface as a function of the spacings $d_2$ and $z_3$
obtained by the density constrained minimization procedure. The minimum
energy is obtained for $d_2=2.65$~fm and $z_3=2.3$~fm. This numerically obtained minimum
corresponds to an equilateral triangle placement of alpha particles, and is identified as the ground state.
These findings are in agreement with other cluster model calculations (see for
example~\cite{yoshida2016}).
In Fig.~\ref{fig:gs}(a) we plot the ground state density as well as the localization function for the
ground state of \textsuperscript{12}C in the $x$-$z$ plane.
The density has a triangular shape and looks relatively compact with an octupole deformation.
Experimentally deduced mass radius of $^{12}$C is 2.43~fm~\cite{tanihata1985,angeli2013}.
Different cluster model calculations yield a range of $2.40-2.53$~fm~\cite{kanada2007}. Our calculations
result in a slightly larger radius of 2.57~fm.
The quadrupole deformation for $^{12}$C is experimentally deduced to be oblate
with $\beta_2=-0.4$~\cite{yasue1983,kelley2017}. Cluster calculations of Ref.~\cite{ichikawa2022}
found $\beta_2=-0.41$ and $\gamma=27.5^{\circ}$. Our calculations find $\beta_2=-0.42$ and $\gamma=29.7^{\circ}$.

Figure~\ref{fig:gs}(b) shows the $n\uparrow$ localization function immediately after placing the three
alphas in their appropriate locations but before the start of the density constraint iterations.
As we have mentioned above, for a single alpha
particle the localization has a fixed value of 1.0 throughout.
Thus the mere combination of three alphas does show a significant localization.
However, the dominant localization is still 1.0 suggesting a pure alpha makeup.
The density constraint minimization modifies this localization function as shown in Fig.~\ref{fig:gs}(c).
The alpha substructure of the ground state is still clearly
pronounced. The regions close to the value 1.0 indicate
the prominent positions of the three alpha particles. It is clear that the alpha particles are
connected by bond like arms. The localization function for protons and spin-down components
essentially show the same structure.

We have previously shown that enforcing the Pauli exclusion principle in density-constraint HF
calculations
has a strong impact on the spin-orbit energy~\cite{simenel2017}, absorbing
 a large part of the Pauli repulsion. This is in agreement with the observations that the spin-orbit
interaction is the primary driver in partially dissolving the alpha clusters in the ground state of
\textsuperscript{12}C~\cite{itagaki2004b,fukuoka2013,horiuchi2023}.

What is also interesting is the evolution of the single-particle parities
during the density constrained minimization procedure.
Initially, all the alpha particles are naturally in
their $s$-states. While we do not have good parity for the deformed state, at the end of the minimization four of the six neutron or proton single-particle
states acquire average negative parity values (not unity), which is appropriate for the ground state of \textsuperscript{12}C.
This has been previously observed in the dynamical collapse of the metastable linear-chain state in
TDHF calculations~\cite{umar2010c}. The conclusion is that within our approach the
ground state of \textsuperscript{12}C is not a pure alpha condensate but more of a molecular
type state formed by the bonding of three alphas.
\begin{figure}[!htb]
    \includegraphics[width=8.6cm]{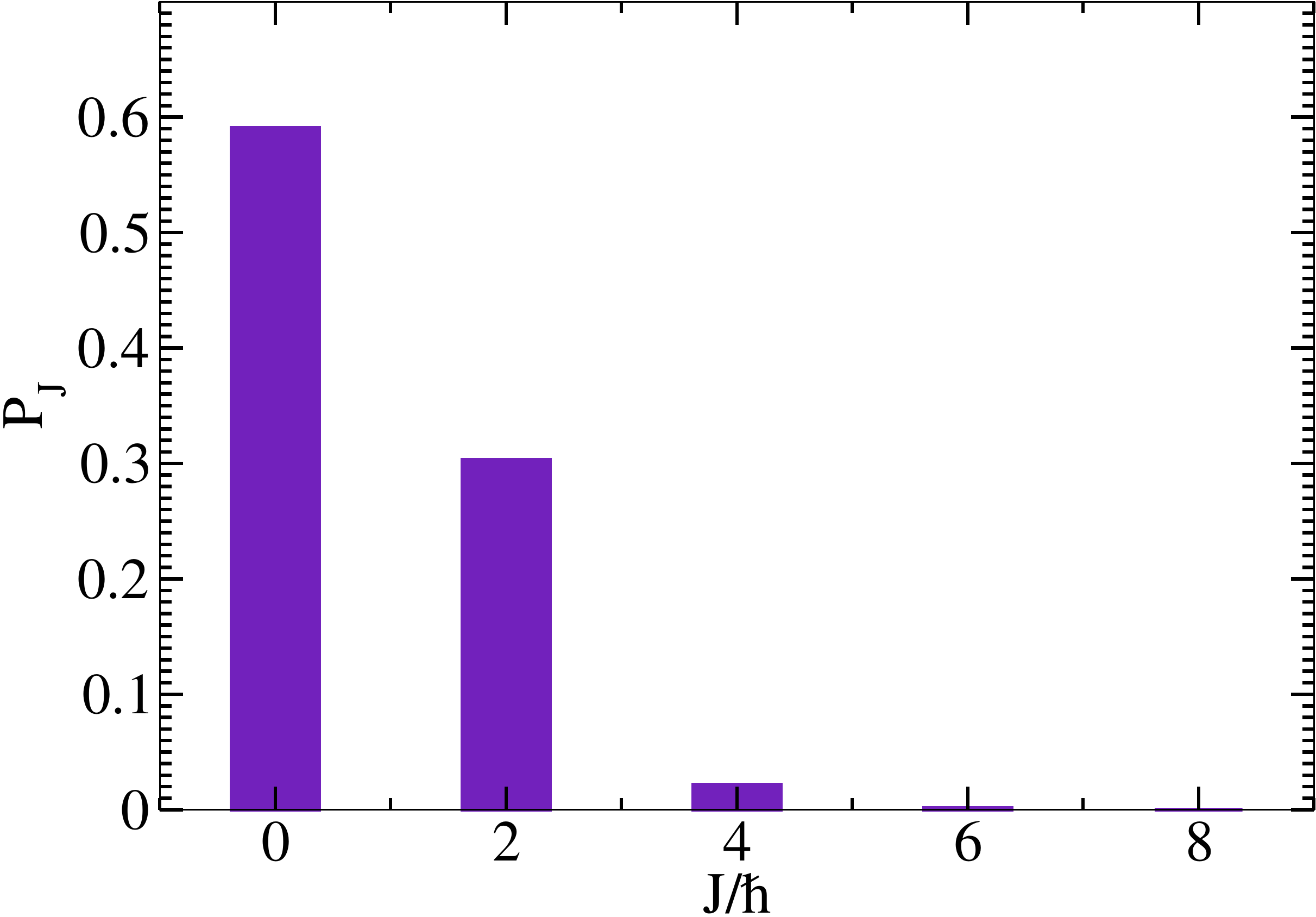}
    \caption{\protect Angular momentum projection of the \textsuperscript{12}C ground state
    configuration.}
    \label{fig:proj}
\end{figure}

We have performed angular momentum projection of this ground state configuration following the
method discussed in Ref.~\cite{shinohara2006}, which is shown in Fig.~\ref{fig:proj}.
It is interesting to see that the major component is $J=0$, as one would expect for the ground state. There is, however, a significant $J=2$ component implying that, in principle, ground state observables (e.g., binding energy) should be evaluated from the $J=0$ projected state. Note that there is little contribution from $J>2$.

Using the same procedure we have also tried to identify the configuration that was observed in
Ref.~\cite{umar2010c}, which could be the candidate for the Hoyle state at DFT level. It
is believed to arise from the bending of the linear-chain three alpha configuration,
which was seen in TDHF calculations of the triple-alpha reaction~\cite{umar2010c}
as an intermediate state during the dynamical collapse of the linear-chain state to the spherical
ground state. There, the dynamical transition of some of the initial single-particle parities from an
$s$-state to a $p$-state was also noted.
\begin{figure}[!htb]
    \includegraphics[width=8.4cm]{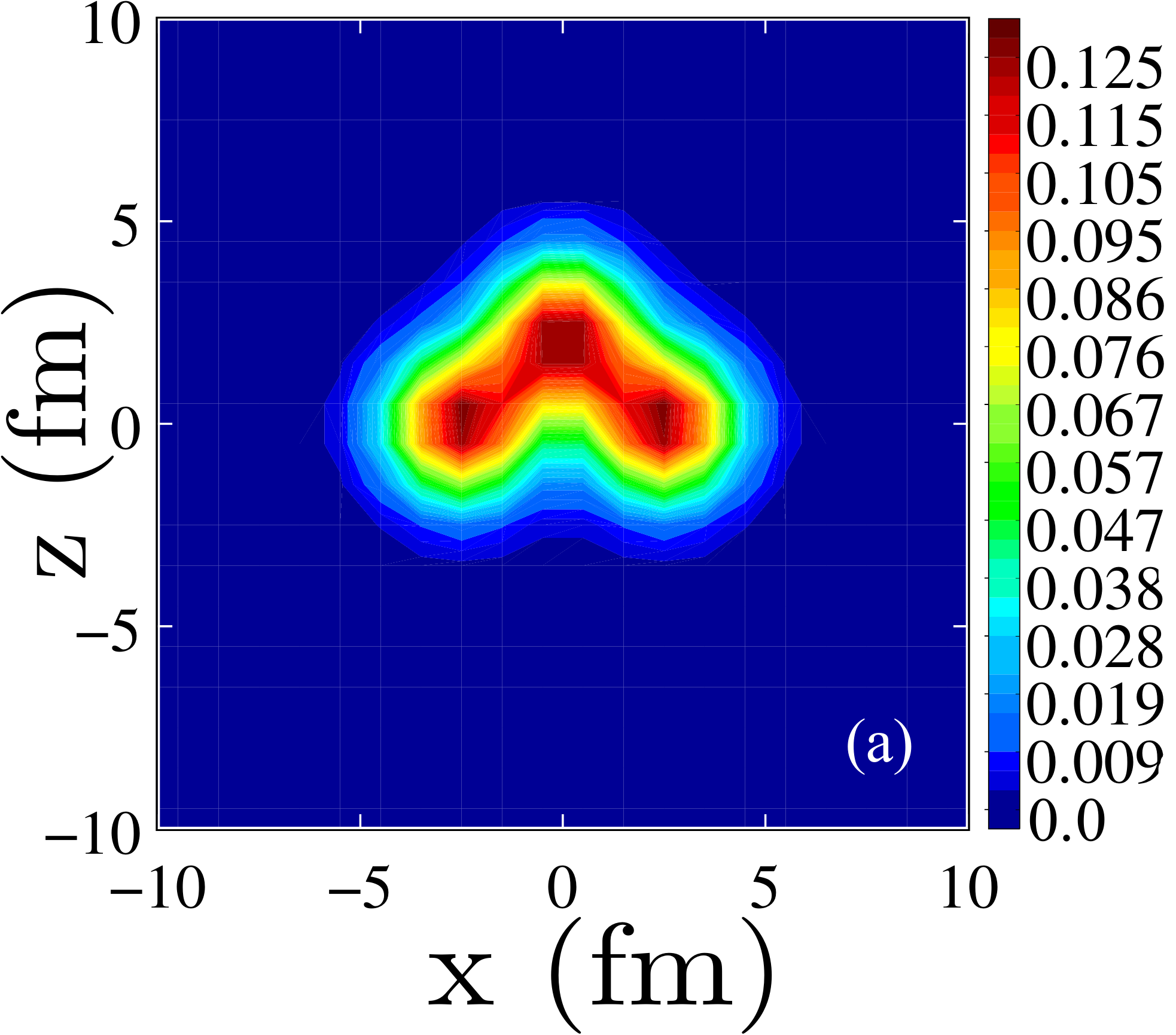}
    \includegraphics[width=8.4cm]{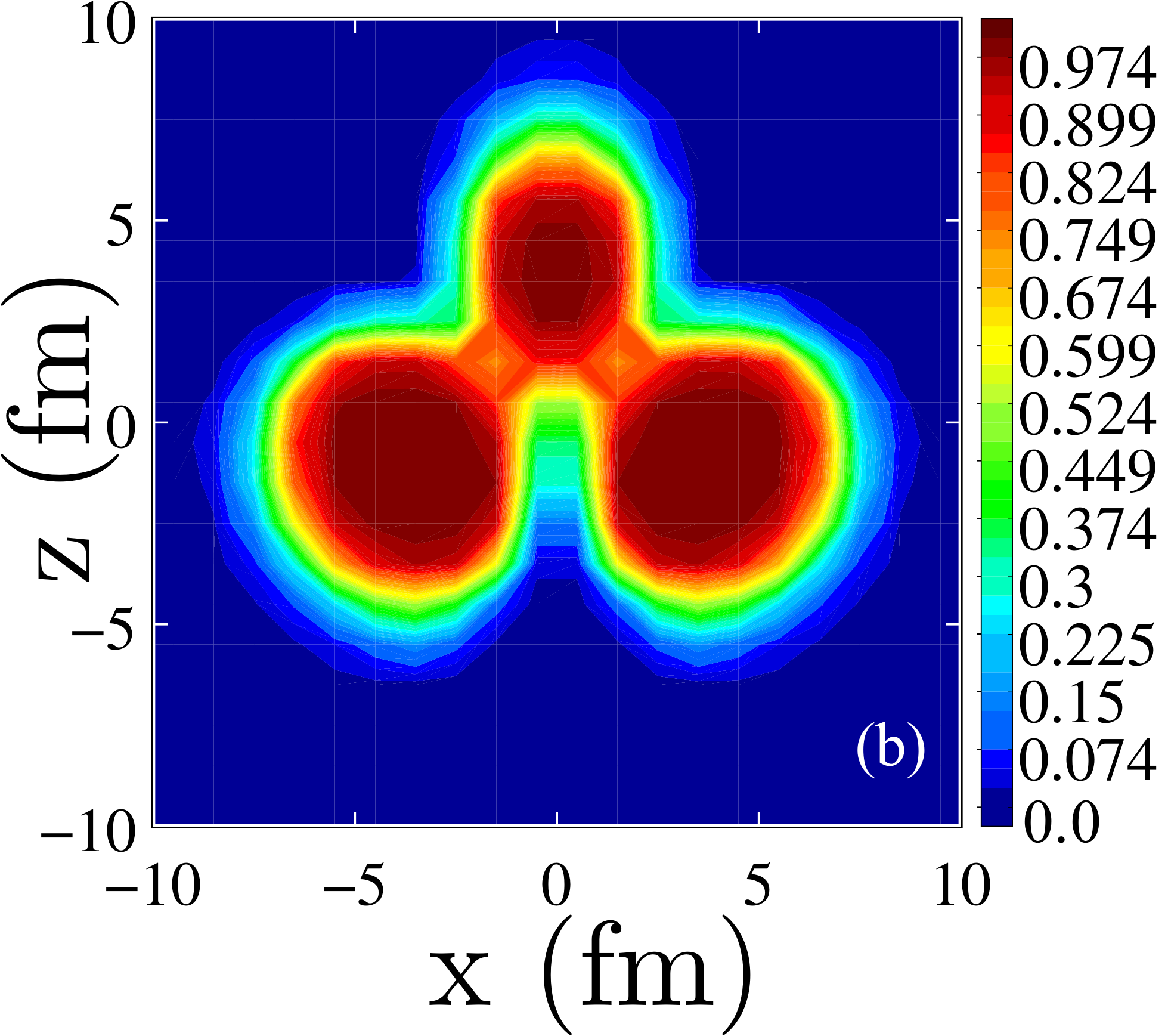}
    \caption{\protect (a) The total density for the bent-arm state configuration of the three alpha particles plotted
        in the $x$-$z$ plane. (b) The corresponding localization function of the bent-arm state configuration. }
    \label{fig:ex2}
\end{figure}
The observed metastable bent-arm configuration occurred during this parity transition.
Here, we also looked at the changing parities of the single-particle states as we changed
the values of $d_2$ and $z_3$. Again, these are not parity projected states so this is simply
a signature for changing single-particle symmetries.
Dependence on parity was also studied in cluster model calculations~\cite{itagaki2006}.
The location of this configuration is shown in Fig.~\ref{fig:PES} with a "+" sign.
As we see PES is very soft in the "d2" direction and this point is a very shallow
local minimum. It is interesting that the dynamical collapse of the linear
chain state also showed that the system spent some time at the bent-arm configuration
but there was no discernible minimum in the DC-TDHF potentials (see Fig.~3 of Ref.~\cite{umar2010c}).
This makes the identification of this configuration more tenuous.
The lowest energy configuration corresponding to this intermediate state is depicted in Fig.~\ref{fig:ex2}
and corresponds to $d_2=5.0$~fm and $z_3=2.2$~fm.
Similar to the ground state case we find this bent-arm mode to be a hybrid configuration of three alphas
with molecular like bonds between the center alpha particle and the ones on each end, as shown in
Fig.~\ref{fig:ex2}(a). Unlike the ground state configuration this configuration has only two main bonds
and has the shape of an obtuse isosceles triangle. The overlap between the Slater
determinant of the ground state configuration and the bent-arm state is on the order of $10^{-3}$,
which is small enough to consider that these are different eigenstates of the system.
The localization function for the bent-arm configuration, shown in Fig.~\ref{fig:ex2}(b), is very
telling. We see that the two clusters
that are on each end are associated with extended $C\sim1$ regions, indicating that they are closer to becoming pure alpha
particles.

\section{Conclusions}\label{sec3}
We have introduced a new framework for studying clusterization in light nuclei, which is based on
the constrained density functional theory.
The new approach does not make any assumptions about the mathematical form of the
single-particle wavefunctions and employs the full effective interaction.
Results show that the \textsuperscript{12}C ground state is an equilateral triangle, which has
a molecular type configuration. The nuclear localization function shows bond like structures
being formed among the original alpha particles as a result of antisymmetrization and energy
minimization.  One can conclude
that these configurations are a hybrid between pure mean-field and a pure alpha particle
condensate.
From our investigation of the cluster energy surface it is clear that a pure alpha condensate
(characterized by pure $s$-wave states) would only occur if the three alphas are relatively far
from each other.

One disadvantage of not using Gaussian type single-particle states or alpha
particles with custom cluster potentials is that we are unable to correct for the spurious
center of mass energy. Another
is that procedures like angular momentum projection, generator coordinate method, etc.
become numerically very challenging for the full effective interaction. This makes detailed
spectroscopic comparisons with experiment very difficult. On the other hand one advantage is that
this ground state of \textsuperscript{12}C may be suitable for fusion barrier calculations
using frozen Hartree-Fock or density constrained frozen Hartree-Fock
methods~\cite{simenel2017,umar2021}, which we plan to investigate in the future.
The preparation of the alpha clustering configuration can also be used for the development and quantification of new energy density functionals, particularly in sectors where the static properties are under-informed by the typical data used in calibration~\cite{mcdonnell2015,godbey2022}.\\

\begin{acknowledgments}
One of the authors (ASU) would like to thank the organizers of the MCD2022 workshop, where some
of ideas presented here have been inspired.
This work has been supported by the U.S. Department of Energy under award numbers DE-SC0013847 (Vanderbilt University),
DE-SC0013365 (Michigan State University), DE-NA0004074 (NNSA, the Stewardship Science Academic Alliances program),
and by the Australian Research Council Discovery Project (project number DP190100256) funding schemes.
\end{acknowledgments}

\bibliography{VU_bibtex_master.bib}


\end{document}